\documentclass[pra,twocolumn,showpacs,letterpaper,showpacs,superscriptaddress]{revtex4-1}

\usepackage{graphicx,amsmath,amssymb,amsfonts,latexsym,color,dcolumn,bm,epsfig,subfigure}

\newcommand{\mathbfh}[1]{\hat{\mathbf{#1}}}

\renewcommand{\imath}[0]{\mathrm{i}}
\newcommand{\abs}[1]{\left\vert#1\right\vert}

\begin{document}

\title{Failure of local thermal equilibrium in quantum friction}

\author{F. Intravaia}
\affiliation{Max-Born-Institut, 12489 Berlin, Germany}
\author{R. O. Behunin}
\affiliation{Department of Applied Physics, Yale University, New Haven, Connecticut 06511, USA}
\author{C. Henkel}
\affiliation{Institute of Physics and Astronomy, University of Potsdam, Karl-Liebknecht-Str. 24/25, 14476 Potsdam, Germany}
\author{K. Busch}
\affiliation{Max-Born-Institut, 12489 Berlin, Germany}
\affiliation{Humboldt-Universit\"at zu Berlin, Institut f\"ur Physik, AG Theoretische Optik \& Photonik, 12489 Berlin, Germany}
\author{D. A. R. Dalvit}
\affiliation{Theoretical Division, MS B213, Los Alamos National Laboratory, Los Alamos, NM 87545, USA}

\begin{abstract} 
Recent progress in manipulating atomic and condensed matter systems has instigated a surge of interest in non-equilibrium physics, including many-body dynamics of trapped ultracold atoms and ions, near-field radiative heat transfer, and quantum friction. Under most circumstances the complexity of such non-equilibrium systems requires a number of approximations to make theoretical descriptions tractable. In particular, it is often assumed that spatially separated components of a system thermalize with their immediate surroundings, although the global state of the system is out of equilibrium. This powerful assumption reduces the complexity of non-equilibrium systems to the local application of well-founded equilibrium concepts. While this technique appears to be consistent for the description of some phenomena, we show that it fails for quantum friction by underestimating by approximately $80 \%$ the magnitude of the drag force. Our results show that the correlations among components of driven, but steady-state, quantum systems invalidate the assumption of local thermal equilibrium, calling for a critical reexamination of this approach for describing the physics of non-equilibrium systems. 
\end{abstract}

\maketitle

%%%%%%%%%%%%%

In recent years the physics of non-equilibrium systems has attracted a lot of attention 
from different disciplines, such as stochastic thermodynamics and many-body quantum dynamics \cite{Speck06,Seifert08,Rigol08}.
In particular, there has been a renewed interest in non-equilibrium dispersion forces. Better known for  
equilibrium phenomena such as the van der Waals/Casimir-Polder force 
\cite{Casimir48a}  and the Casimir effect \cite{Casimir48}, 
these interactions play an important role in several fields of physics, including atomic  
\cite{Fortagh07}
and statistical physics \cite{Fisher78,Gambassi09a}, gravitation \cite{Onofrio06} and cosmology \cite{Adler95}. 
Non-equilibrium physics enters in the description of these phenomena when, for example, 
temperature gradients or mechanical motion become relevant elements of the system. 

From the theoretical standpoint, one must often rely on approximations in order to predict 
the non-equilibrium physics of a specific system. One of the most ubiquitous 
approaches relies on the local thermal equilibrium (LTE) approximation, which consists in treating 
the individual components of a system as if they were in 
local thermal equilibrium with their immediate surroundings. The main advantage of such a technique is that common equilibrium tools, such as the fluctuation-dissipation theorem (FDT) 
\cite{Callen51}, can  be applied locally, and then these local results are combined to describe the 
non-equilibrium dynamics of the full system. 
The usual justification for the LTE approximation 
is that the correlation length of the fields that mediate the interactions is often rather short (the dynamics in sufficiently 
well-separated locations are incoherent and can be treated as being independent 
\cite{Polder71,Eckhardt84}), and the sub-systems locally relax to equilibrium on a fast time-scale.
The LTE approximation has been used in several non-equilibrium contexts, such as
near-field radiative heat transfer \cite{Polder71}, Casimir forces between bodies at different temperatures  
\cite{Dorofeyev98,Obrecht07}, and quantum friction \cite{Pendry97,Volokitin07,Hoye10,Maghrebi12,Maghrebi13a,Hoye14}.
In these previous cases, however, a quantitative assessment of the LTE approximation 
is missing. 
In this work, we show that this common approach actually fails to provide reliable 
predictions for quantum friction. 

Let us consider an atom moving in vacuum with nonrelativistic velocity $v$ at a distance $z_{a}>0$ above and parallel to a flat surface placed at $z=0$
(see Fig. 1). 
The atom couples to the electromagnetic field via its dipole moment $\hat{\bf d}(t)$.
In previous work \cite{Intravaia14} it was shown that the zero-temperature frictional force (quantum friction) acting on the atom is given by
\begin{multline}
\mathbf{F}_{\rm fric}= - 2\int_{0}^{\infty}d\omega\int\frac{d^{2}\mathbf{k}}{(2\pi)^{2}} \\
\times \mathbf{k} 
\mathrm{Tr}\left[\underline{S}(\mathbf{k}\cdot\mathbf{v}-\omega;\mathbf{v})\cdot \underline{G}_{I}(\mathbf{k},z_{a}, \omega)\right].
\label{friction2t}
\end{multline} 
Here,  $\underline{S}(\omega;\mathbf{v})$ is the non-equilibrium velocity-dependent dipole power spectrum tensor (related to the spectral distribution of energy in the dipole),
and $\underline{G}(\mathbf{k},z_{a}, \omega)$ is the Fourier transform (in time and along the $(x, y)$ plane) of the Green tensor describing the electromagnetic response of the surface.
In the following the subscript $I$ ($R$) means that the imaginary (real) part has to be (component-wise) considered. 

\begin{figure}[t]
\includegraphics[width=8.5cm]{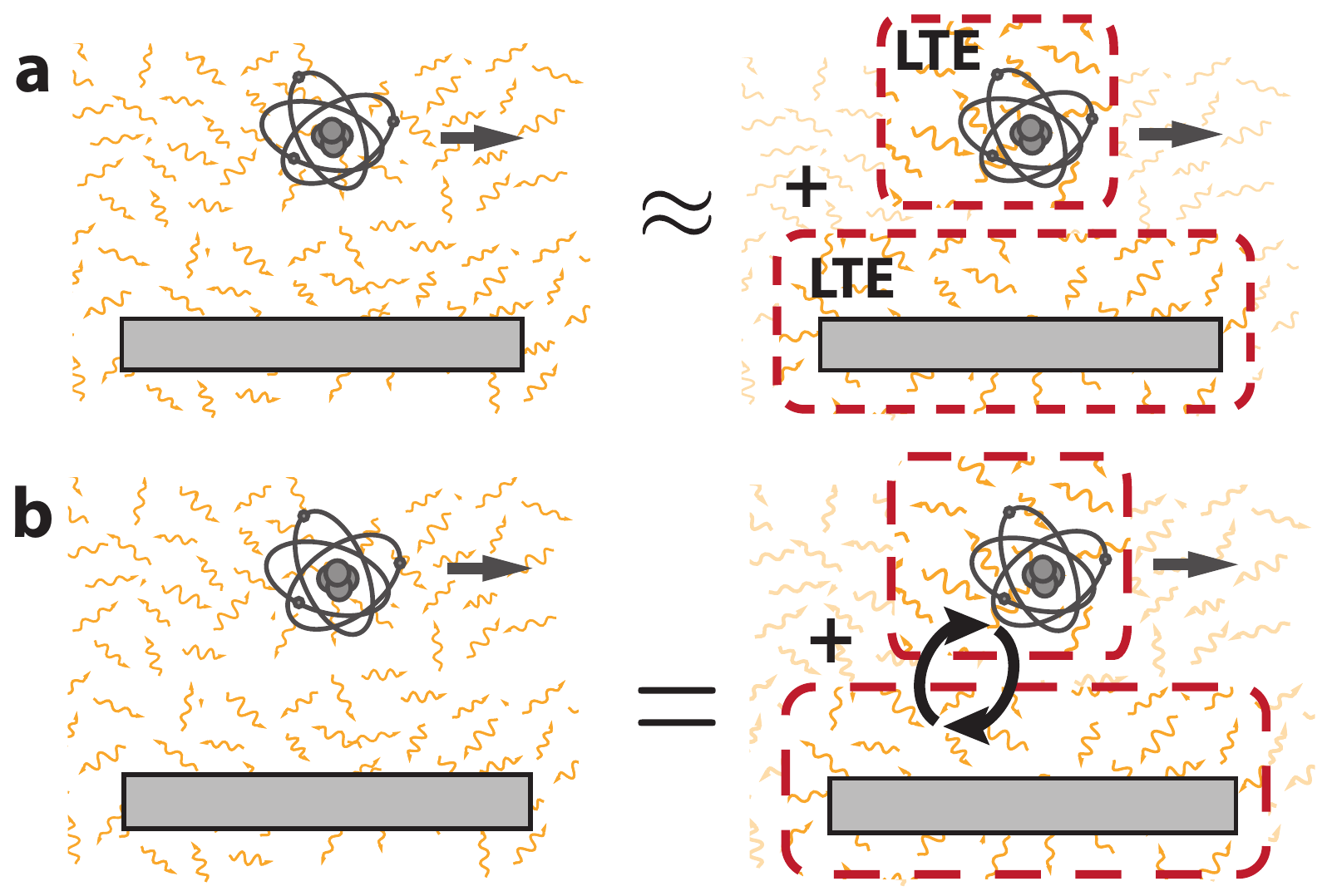}
\caption{%{\color{blue} Remove FTE and their arrows. Make all fonts, lines, etc, consistent|}
Schematic representation of the difference between the LTE approximation (a)
and the full non-equilibrium description (b) for quantum friction. In the first case it is assumed that the atom and
the surface are separately in thermal equilibrium with their immediate local environments. This
description applies the fluctuation-dissipation theorem for each sub-system, to approximatively describe the full non-equilibrium system. 
Correlations between the atom and surface (pictorially represented by the black arrows in (b)) lead to a failure of the LTE approximation, which 
underestimates the magnitude of quantum friction by approximately $80 \%$ (see the main text).
}
\label{LTEfig}
\end{figure}

The standard approach used in the literature to compute the frictional force has been to resort to the LTE approximation. It is assumed that the particle and the 
surface surrounded by its electromagnetic field are \emph{locally} at thermal equilibrium at $T=0$
in their respective rest frames, 
and that the fluctuation-dissipation theorem separately applies to each subsystem \cite{Hoye10,Maghrebi12,Maghrebi13a,Hoye14}. In this case one assumes that
$\underline{S}(\omega;\mathbf{v})$ is related to the imaginary part of the particle's polarizability tensor $\underline{\alpha}(\omega;\mathbf{v})$ via the zero-temperature FDT,
\begin{equation}
 \underline{S}(\omega; \mathbf{v})\approx\frac{\hbar}{\pi} \theta(\omega)\underline{\alpha}_{I}(\omega;\mathbf{v}).
\label{eqFDT}
\end{equation}
(the function $\theta(\omega)$ is the Heaviside function). Upon implementing the LTE the resulting frictional force at low velocities takes the form \cite{Intravaia14}
\begin{equation}
F_{\rm fric} \approx  -\frac{2\hbar v^{3}}{3(2\pi)^{3}}
\int_{-\infty}^{\infty}\hspace{-.2cm}dk_{y}\int_{0}^{\infty}\hspace{-.2cm}dk_{x}\, k_{x}^{4} \mathrm{Tr}\left[\underline{\alpha}'_{I}(0)\cdot \underline{G}'_{I}(\mathbf{k},z_{a}, 0)\right],
\label{lowvelocity}
\end{equation}
where the primes denote frequency derivatives and we assumed that the motion is along the $x$-direction. 
A detailed quantitative evaluation of equation \eqref{lowvelocity} requires the 
low-frequency behavior of the polarizability, which is often calculated within second-order perturbation theory \cite{Milonni04,Buhmann04,Lach12,Jentschura15}.
Although the use of the LTE approximation can be justified within a second-order perturbative approach in the dipole strength for particles with large intrinsic dissipation \cite{Intravaia16}, 
it becomes less rigorous for atoms where dissipation is induced by the interaction with the electromagnetic field. 
For systems where dissipation is caused by radiative damping, quantum friction requires a higher-than-second-order perturbative calculation, and hence
the local thermal equilibrium approximation fails because the description of such systems necessarily encompasses the correlations between the atom and the surface
(see Fig. \ref{LTEfig}). This is the key insight of this paper.

%\section*{\large Results}

In order to test the validity of the LTE approximation in quantum friction, we are going to compute the dipole power spectrum $\underline{S}(\omega;\mathbf{v})$, evaluate the resulting drag force, and compare it to the LTE result. This entails the computation of the non-equilibrium steady-state (NESS) of the joint atom+field+matter system. This difficult problem becomes manageable by modeling the internal atomic dynamics as a harmonic oscillator \cite{Einstein10}, 
for which it is possible to obtain an exact, non-perturbative form for the dipole power spectrum thanks to the quadratic nature of the full system Hamiltonian \cite{Intravaia14}.
We work in the Heisenberg picture to calculate the dipole correlator in the steady state and derive the power spectrum.

%%%%
%%%%%
 
In the non-relativistic approximation 
the equation of motion of a dipole moving along a prescribed trajectory 
$\mathbf{r}_{a}(t)$ and with a fixed direction ${\bf d}$ is given by 
\begin{equation}
\ddot{\mathbfh{d}}(t)+\omega_{a}^{2}\mathbfh{d}(t)=\frac{2\omega_{a}}{\hbar}\mathbf{d}\mathbf{d}\cdot \mathbfh{E}(\mathbf{r}_a(t),t), 
\label{eqnMotionOsc}
\end{equation}
where $\omega_{a}$ is the oscillator's frequency and $\mathbfh{E}$ is the electric field. We assume that the oscillator has no intrinsic dissipation - all dissipative dynamics arises 
from the coupling to the electromagnetic field. The electric field at the instantaneous position 
of the atom is given by
\begin{widetext}
\begin{equation}
\mathbfh{E}(\mathbf{r}_a(t),t)=\mathbfh{E}_{0}(\mathbf{r}_a(t),t)
+\int_{t_{i}}^{t}dt' \int_{-\infty}^{\infty}d\omega \,e^{-\imath \omega(t-t')}
\int \frac{d^{2}\mathbf{k}}{(2\pi)^{2}}\underline{G}(\mathbf{k},z_{a}, \omega)
e^{\imath\mathbf{k}\cdot(\mathbf{R}_a(t)-\mathbf{R}_a(t'))}\cdot\mathbfh{d}(t'),
\label{emfield2}
\end{equation}
\end{widetext}
where $\mathbfh{E}_{0}$ is the field that is generated by the quantum fluctuating currents 
in the medium. In the stationary $t \rightarrow \infty$ limit, we use that ${\bf R}_a(t)={\bf R}_a + {\bf v} t$ 
and ${\bf R}_a(t')={\bf R}_a + {\bf v} t'$. Upon inserting \eqref{emfield2} into the equation 
of motion for the dipole, we obtain the stationary solution in Fourier space as
\begin{equation}
\mathbfh{d}(\omega) =
\int \frac{d^{2}\mathbf{k}}{(2\pi)^{2}} \, \underline{\alpha}(\omega;\mathbf{v}) \cdot 
\mathbfh{E}_{0}(\mathbf{k},z_{a}, \omega+\mathbf{k}\cdot\mathbf{v})
e^{\imath\mathbf{k}\cdot\mathbf{R}_{a}}~,
\label{d}
\end{equation}																						
where we have defined the velocity-dependent polarizability
\begin{equation}
\underline{\alpha}(\omega;\mathbf{v})=\frac{2\omega_{a}}{\hbar}
\frac{\mathbf{d}\mathbf{d}}{ \omega_{a}^2  -\Delta(\omega;\mathbf{v})-\omega^{2}-\imath \omega \gamma(\omega;\mathbf{v})} ~.
\label{alpha}
\end{equation}
In this expression, $\gamma$ is the radiative damping while $\Delta$ is related to a frequency shift \cite{Intravaia14,Klatt16},
and they are given by 
\begin{subequations}
\begin{equation}
\Delta (\omega;\mathbf{v})=\frac{2 \omega_a}{\hbar}\int \frac{d^{2}\mathbf{k}}{(2\pi)^{2}}\mathbf{d}\cdot\underline{G}_{R}(\mathbf{k},z_{a}, \omega+\mathbf{k}\cdot\mathbf{v})\cdot\mathbf{d},
\label{cpshift}
\end{equation}
\begin{equation}
\gamma(\omega; \mathbf{v})=\frac{2\omega_{a}}{\hbar\omega}\int \frac{d^{2}\mathbf{k}}{(2\pi)^{2}}\mathbf{d}\cdot\underline{G}_{I}(\mathbf{k},z_{a}, \omega+\mathbf{k}\cdot\mathbf{v})\cdot\mathbf{d} .
\label{gammaind}
\end{equation}
\end{subequations}

The dipole correlation function, derived from equation \eqref{d}, defines the power spectrum
$\langle\mathbfh{d}(\omega)\mathbfh{d}(\omega')\rangle = (2\pi)^{2}\underline{S}(\omega;\mathbf{v}) \delta(\omega+\omega')$,
where the average is taken over the initial factorized state of the system, 
$\hat{\rho}(t_{i})=\hat{\rho}_a(t_{i})\bigotimes\hat{\rho}_{\rm f/m}(t_{i})$. Here, 
$\hat{\rho}_a(t_{i})$ is the atom's initial density matrix and $\hat{\rho}_{\rm f/m}(t_{i})$ 
represents the state of the coupled field plus matter subsystem. Both the atom and the field+matter are assumed to be 
initially in their respective ground states. Because of equation \eqref{d}, we can compute the dipole-dipole 
correlation in terms of the field-field correlator. Since $\mathbfh{E}_{0}$ is the field generated solely by the surface we can use the FDT. 
This gives
\begin{multline}
\langle \mathbfh{E}_0(\mathbf{k},z_{a},\omega)\mathbfh{E}_0(\mathbf{k}',z_{a},\omega')\rangle\\
 = 2(2\pi)^{3}\hbar\theta(\omega) \underline{G}_{\Im}(\mathbf{k},z_{a},\omega)\delta(\omega+\omega')\delta(\mathbf{k}+\mathbf{k}')~,
 \end{multline}
where we have defined
$\underline{G}_{\Im}(\mathbf{k},z,\omega)
 =
 [\underline{G}(\mathbf{k},z,\omega)-\underline{G}^{\dag}(\mathbf{k},z,\omega)]/(2 \imath)$. By combining all the above equations and using that $\underline{\alpha}(-\omega;\mathbf{v})=\underline{\alpha}^{*}(\omega;\mathbf{v})$, 
we obtain 
\begin{multline}
\underline{S}(\omega;\mathbf{v}) = \frac{\hbar}{\pi}\int \frac{d^{2}\mathbf{k}}{(2\pi)^{2}}\,\theta(\omega+\mathbf{k}\cdot\mathbf{v}) \\
\times\underline{\alpha}(\omega;\mathbf{v})\cdot
\underline{G}_{I}(\mathbf{k},z_{a},\omega+\mathbf{k}\cdot\mathbf{v})\cdot\underline{\alpha}^{*}(\omega;\mathbf{v}) .
\label{spectrum}
\end{multline}
Since the matrix $\mathbf{d}\mathbf{d}$ is a symmetric tensor, we replaced $\mathbf{d}\cdot  \underline{G}_{\Im}(\mathbf{k},z_{a},\omega)\cdot\mathbf{d}$
by $\mathbf{d}\cdot\underline{G}_{I}(\mathbf{k},z_{a},\omega)\cdot\mathbf{d}$, where only the symmetric part of $\underline{G}_I$ contributes to the tensor product \cite{Intravaia16}.
By noting that the polarizability and the Green tensor are related via
\begin{equation}
\underline{\alpha}_{I}(\omega;\mathbf{v})=\int \frac{d^2\mathbf{k}}{(2\pi)^{2}}\underline{\alpha}(\omega;\mathbf{v})\cdot \underline{G}_{I}(\mathbf{k},z,\omega+\mathbf{k}\cdot\mathbf{v})\cdot \underline{\alpha}^{*}(\omega;\mathbf{v})~,
\end{equation}
the dynamic power spectrum can be expressed as
\begin{equation}
\underline{S}(\omega;\mathbf{v})=\frac{\hbar}{\pi} \theta(\omega)\underline{\alpha}_{I}(\omega;\mathbf{v})+\frac{\hbar}{\pi}\underline{J}(\omega;\mathbf{v}),
\label{NEFDT}
\end{equation}
where
\begin{multline}
\underline{J}(\omega;\mathbf{v})=\int \frac{d^{2}\mathbf{k}}{(2\pi)^{2}}\left[\theta(\omega+\mathbf{k}\cdot\mathbf{v})-\theta(\omega)\right] \\
\times\underline{\alpha}(\omega;\mathbf{v})\cdot
\underline{G}_{I}(\mathbf{k},z_{a},\omega+\mathbf{k}\cdot\mathbf{v})\cdot\underline{\alpha}^{*}(\omega;\mathbf{v}).
\end{multline}
Equation \eqref{NEFDT} constitutes the \emph{generalized non-equilibrium FDT} for the moving 
harmonic oscillator. It shows that, when the system is in a NESS, an extra term $\underline{J}$ is added to the standard FDT, equation \eqref{eqFDT}.
The expression in \eqref{NEFDT} is similar to classical non-equilibrium generalizations of the FDT (see, for example, Refs.\cite{Agarwal72,Chetrite08,Baiesi09,Talkner09,Seifert10}), where 
the additional term is related to entropy production.
However, these works often include assumptions (e.g. Markovianity) which are incompatible with the description of quantum friction \cite{Intravaia16}.

Upon inserting equation \eqref{NEFDT} into \eqref{friction2t}, we obtain two distinct contributions 
to the quantum frictional force, 
\begin{equation}
\mathbf{F}_{\rm fric} = \mathbf{F}^{\rm LTE}_{\rm fric} + \mathbf{F}^{J}_{\rm fric},
\label{alphaJ}
\end{equation}
which respectively arise from the first and second terms on the right hand side of \eqref{NEFDT}. 
As we will show below, the low-velocity expansion of $\mathbf{F}^{\rm LTE}_{\rm fric}$ corresponds to equation \eqref{lowvelocity}
\cite{Volokitin07,Hoye14,Dedkov03},  while $\mathbf{F}^{J}_{\rm fric}$ is entirely due to the non-equilibrium dynamics of 
our system. We now compute 
the low-velocity expansion of the force in equation \eqref{alphaJ}. 
As before, we assume that the motion occurs along the $x$-direction, so that $\mathbf{F}_{\rm fric} =F_{\rm fric} \,\mathbf{x}$ 
(here $\mathbf{x}$ is the unit vector along the $x$ direction). The total Green tensor in equation\eqref{friction2t} can be decomposed 
as the sum of the vacuum $\underline{G}_{0}$ and the scattered contribution $\underline{g}$. 
Because of Lorentz invariance, the vacuum contribution $ \underline{G}_{0}$ does not contribute to the frictional force \cite{Kyasov02,Dedkov03,Volokitin08,Pieplow13}. For simplicity, we consider the near-field limit for $\underline{g}$, whose symmetric part has an imaginary part given by \cite{Wylie84}
\begin{equation}
\underline{g}_{I}(\mathbf{k},z_{a};\omega) = \frac{r_{I}(\omega)}{2\epsilon_{0}}ke^{-2k z_{a}} \left(\frac{k_{x}^{2}}{k^{2}} \mathbf{x} \mathbf{x}+\frac{k_{y}^{2}}{k^{2}} \mathbf{y} \mathbf{y}+  \mathbf{z} \mathbf{z} \right) ,
\label{nearfieldG}
\end{equation}
where $k=|\mathbf{k}|=\sqrt{k_{x}^{2}+k_{y}^{2}}$, $\epsilon_{0}$ is the vacuum permittivity, and 
$r(\omega)$ is the quasi-static approximation of the transverse magnetic reflection coefficient for the planar surface.  
Using the previous expression one can show that that in the low-velocity limit the first term in the right hand side of equation \eqref{alphaJ} gives (see Supplemental Material) 
\begin{equation}
\bar{F}^{\rm LTE}_{\rm fric} \approx -\frac{90 \bar{\mathcal{A}}^{\rm LTE}}{\pi^{3} }\hbar
\alpha_{0}^{2} \rho^{2} \frac{v^{3}}{(2z_a)^{10}}~,
\label{fricalpha}
\end{equation}
where $\alpha_{0}=2|\mathbf{d}|^{2}/(3\hbar \omega_{a})$ is the static isotropic atomic polarizability, $\rho$ is the material resistivity, and $\bar{\mathcal{A}}^{\rm LTE}=21/20 \approx 1$ is a geometrical factor coming from the average over all dipole orientations.
The above expression reduces to equation \eqref{lowvelocity} when we use the polarizability given in \eqref{alpha}. 
For the non-equilibrium correction term in \eqref{alphaJ}, we obtain similarly (see Supplemental Material)
\begin{equation}
\bar{F}^{J}_{\rm fric} \approx -\frac{72 \bar{\mathcal{A}}^{J}}{\pi^{3} }\hbar
\alpha_{0}^{2} \rho^{2} \frac{v^{3}}{(2z_a)^{10}}~,
\label{fricj}
\end{equation}
where $\bar{\mathcal{A}}^{J}=87/80$. Adding the low-velocity expansions of $\bar{F}^{\rm LTE}_{\rm fric}$ and $\bar{F}^{J}_{\rm fric}$, the full quantum frictional force becomes
\begin{equation}
\bar{F}_{\rm fric} \approx -\frac{864}{5\pi^{3} }\hbar
\alpha_{0}^{2} \rho^{2} \frac{v^{3}}{(2z_a)^{10}}~,
\label{fricTot}
\end{equation}
which differs by almost a factor of two from the approximate LTE result in equation \eqref{fricalpha}. 
This is the main result of our paper and demonstrates that the non-equilibrium contribution to the frictional force is certainly \emph{not} negligible.

%%%%%%%%%%%%%%

%\section*{\large Discussion}

\begin{figure}[t]
\includegraphics[width=8.5cm]{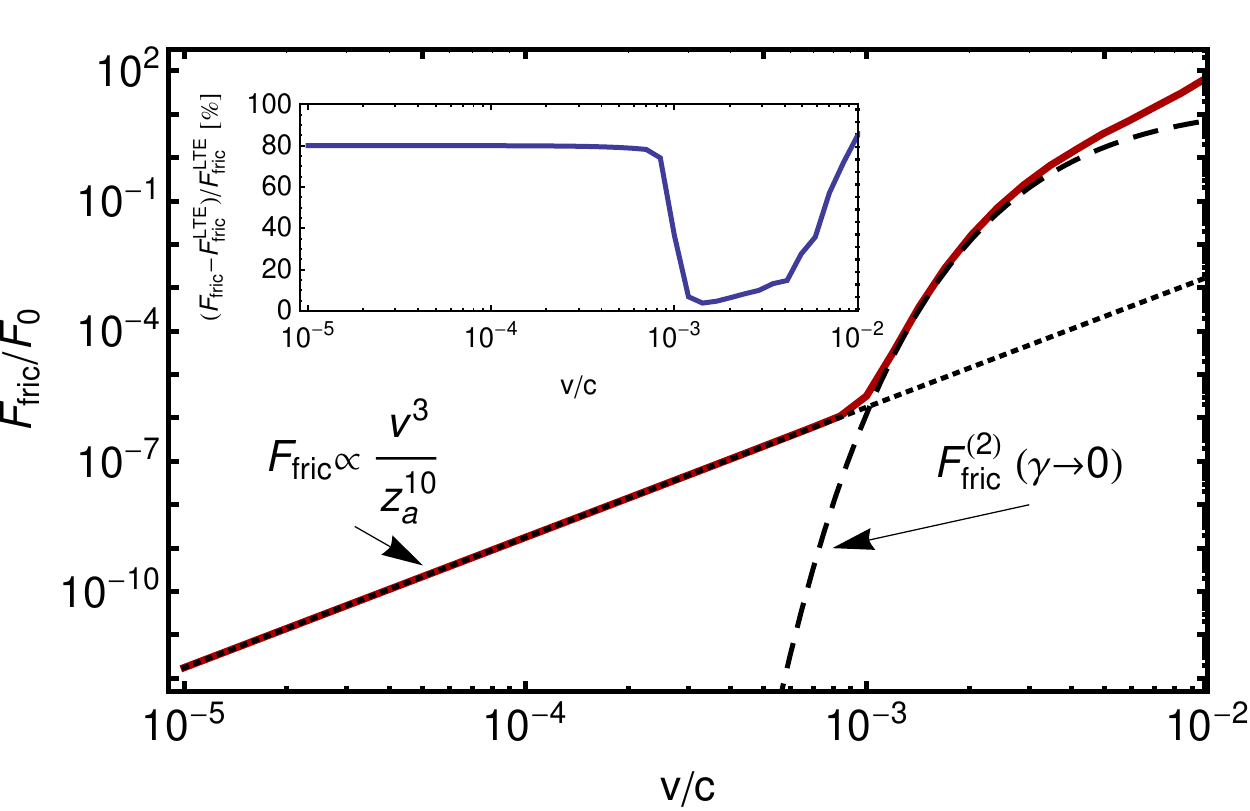}
\caption{
%{\color{blue} In the inset, write $(F_{\rm fric} -F_{\rm fric}^{\rm LTE})/ F_{\rm fric}^{\rm LTE} (\%)$.
%For the dashed black line, I would write $F_{\rm fric}^{(2)}$ instead of the thing with the $\gamma$.}
Velocity dependence of the (normalized) quantum friction of a harmonic oscillator that moves with velocity $v$ 
above a metallic surface described by the Drude model. 
The dipole is oriented  along the direction $(1/\sqrt{3},1/\sqrt{3},1/\sqrt{3})$. 
The oscillator has a resonance frequency $\omega_{a}/\omega_{\rm sp}=0.2$ and moves at a distance 
$z_{a}\omega_{\rm sp}/c=10^{-1}$ above and parallel to the surface with dissipation rate $\Gamma/\omega_{\rm sp}= 0.1$. 
At low velocities the LTE approximation underestimates the frictional force by approximately  $80 \%$ (see inset). As the velocity increases ($v/c\gtrsim 10^{-3}$ for the parameters 
above), the oscillator's radiative damping becomes less relevant and the force is  accurately 
described by the asymptotic expression given in equation \eqref{asympt2} (dashed black line), which corresponds to $\gamma \rightarrow 0$. 
A further increase in the velocity enhances the non-equilibrium contribution to the force and a deviation from the LTE description occurs again (see inset).
The normalization is $F_{0}= -3\hbar \omega_{\rm sp}^5 \alpha_{0} / (2 \pi \epsilon_{0}c^4)$.  For a $^{87} {\rm Rb}$ atom ($\alpha_0=5.26 \times 10^{-39} \; {\rm F} {\rm m}^2$ \cite{Steck08}) and a plasma frequency $\omega_p=9$ eV, we have   $F_0=0.31 \;{\rm fN}$.
}
\label{frictionFull}
\end{figure}

In Fig. \ref{frictionFull} we depict the quantum frictional force equation \eqref{friction2t} 
as a function of velocity. For simplicity the dipole is oriented along a direction for which $\mathcal{A}^{\rm LTE}=\mathcal{A}^{J}=1$ and moving above a metallic surface described by the Drude model $\epsilon(\omega)=1-\omega_{p}^{2}[\omega(\omega+\imath \Gamma)]^{-1}$, where $\omega_p$ is the plasma frequency and $\Gamma$ is the metal's relaxation rate (in this case the material resistivity is given by $\rho=\Gamma/(\epsilon_{0}\omega_{p}^{2})$).
For small velocities the friction is well described by the asymptotic expression equation \eqref{fricTot} (black dotted line). In this region the integrals in equation \eqref{friction2t} are dominated by the low frequency behavior of $\underline{S}(\omega;\mathbf{v})$ and $\underline{G}(\mathbf{k},z_{a}, \omega)$, resulting in the power-law dependency on velocity and separation (see Supplemental Material). The relative difference between the exact and LTE results is more than $80\%$ in this region (see inset of  Fig. \ref{frictionFull}).
At high velocities ($v/c\gtrsim 10^{-3}$ for the parameters in Fig. \ref{frictionFull}) a kink is visible at the crossing between the previous asymptotic expressions and
\begin{multline}
F^{(2)}_{\rm fric}\approx -\frac{\hbar \omega_{\rm sp}^4 \alpha_{0}}{\pi c^4}
 \frac{\Gamma}{16 \epsilon_{0}}\\
\times\sqrt{\frac{ \left(\frac{\omega_a}{\omega_{\rm sp}}\right)^{7}}{\pi \left(\frac{\omega_{\rm sp} z_{a}}{c}\right)^{5}\left(\frac{v}{c}\right)^{3}}}
\left(1+\frac{5 v}{2 z_a \omega_a} \right) e^{-\frac{2 z_a \omega_a}{v}}~
\label{asympt2}
\end{multline}
(black dashed curve in Fig. \ref{frictionFull}), where $\omega_{\rm sp}=\omega_{p}/\sqrt{2}$ is the surface plasmon frequency . 
Equation \eqref{asympt2} is the result of a second-order perturbative expansion of \eqref{friction2t}, and can be explained by a resonant process involving the atom-surface interaction  \cite{Intravaia16}. As shown in the figure, the expression in \eqref{asympt2}
describes well the behavior of the quantum frictional force immediately after the kink. Since the impact of radiative damping is negligible in this second-order expansion, in the region right after the kink the atom and the surface can be considered uncorrelated to a good approximation, and the 
LTE description is sufficient to characterize the quantum frictional process. This is clearly seen in the sharp decrease of the relative difference between the exact and LTE results (see inset). A further increase of the velocity leads again to a deviation from the LTE approximation due to the strengthening of the non-equilibrium-induced atom-surface correlations.
 
 %------------------
%------------------

Due to its small value, an experimental detection of quantum friction is challenging and designing setups that increase 
the strength of the interaction is certainly desirable. 
%Our analysis indicates that the atomic and  material properties are relevant parameters to focus on.  
Specifically, equation \eqref{fricTot} can be rewritten as
\begin{equation}
\bar{F}_{\rm fric} \approx -\frac{216}{5\pi}\hbar
\gamma^{2}(z_{a}) \frac{v^{3}}{(2z_a \omega_{a})^{4}}~,
\end{equation}
where $\gamma(z_{a})=\alpha_{0}\omega_{a}^{2}\rho/(4\pi z_a^{3})$  is the 
leading-order (i.e., low-frequency and small-velocity) expansion of the function $\gamma(\omega; \mathbf{v})$ 
defined in equation \eqref{gammaind}.
This demonstrates that, at low velocities, quantum friction is proportional to the square of the induced decay rate. This feature suggests possible pathways to increase the strength of the  quantum frictional force. For instance, material properties or geometric configurations, such as hyperbolic nanostructures \cite{Intravaia15a}, 
which are known for producing large Purcell factors, are potentially favorable for enhancing the quantum frictional force.

%%%%%%%%%%%%%%%
%
%\section*{\large Discussion}

In conclusion, we have shown that the local thermal equilibrium approximation fails in quantum friction.
We demonstrated this point with an exact solution to a model of a harmonic oscillator moving parallel to a surface, in which the LTE approach underestimates the quantum friction force by approximately $80\%$. Motion-induced atom-surface correlations are ultimately responsible to the breakdown of the local equilibrium assumption. It is worth emphasizing that, despite its extensive application and even if quite reasonable in most circumstances, the LTE approximation relies more on phenomenological considerations than on quantitative estimations. Our results in quantum friction call for a critical assessment of the range of applicability of local thermal equilibrium in other non-equilibrium dispersion interactions. Such an analysis could potentially provide new insights and unravel important features of these and other non-equilibrium systems.

%\vspace{1cm}
%\noindent {\bf Acknowledgments.}

{\it Acknowledgments.} We acknowledge support by the LANL LDRD program, and by the Deutsche Forschungsgemeinschaft (DFG) through 
project B10 within the Collaborative Research Center (CRC) 951 Hybrid Inorganic/Organic Systems 
for Opto-Electronics (HIOS). FI further acknowledges financial support from the European Union Marie Curie People program through the Career Integration Grant No. PCIG14- GA-2013-631571. CH and FI acknowledge support from the DFG through the 
DIP program (grant FO 703/2-1 and SCHM 1049/7-1)

%%%%%%%%%%%%%%%%%%%%%%%%%%%%%%%%%%%%%%%%%%%%%%
%\bibliographystyle{/Users/nabu/Documents/Mydocs/Lavoro/bibliography/bibstyle/prsty}
%\bibliography{/Users/nabu/Documents/Mydocs/Lavoro/bibliography/biblio}

%%%%%%%%%%%%%%%%%%%%%%%%%%%%%%%%%%%%%%%
%\newpage

\clearpage

\setcounter{page}{1}
\setcounter{figure}{0}
\setcounter{equation}{0}
\renewcommand{\theequation}{S\arabic{equation}}
\renewcommand{\figurename}{\textbf{Supplementary Figure}}
\renewcommand{\thefigure}{{\bf S\arabic{figure}}}

\section*{\large Supplemental Material}

When the motion occurs along the $x$-axis, it is possible to carry out the integration over $k_{y}$ 
in equation \eqref{friction2t} analytically. 
We define
\begin{equation}
g(k_x z_a, \omega)\equiv  \frac{\epsilon_{0} z_a^{2}}{3\alpha_{0}}\int_{-\infty}^{\infty}\frac{dk_{y}}{2\pi} \mathrm{Tr}\left[ 2\frac{\mathbf{d}\mathbf{d}}{\hbar \omega_{a}}\cdot\underline{g}({\bf k},z_{a};\omega)\right] ,
\end{equation}
where $g(w, \omega)=r(\omega)w^2 \mathcal{K}(w,\varphi,\theta)/(6 \pi)$, $r(\omega)$ is the reflection coefficient, and 
\begin{equation}
\mathcal{K}(w,\varphi,\theta)=A_{0}(\varphi,\theta)K_{0}(2\abs{w})+A_{2}(\varphi,\theta)K_{2}(2\abs{w}).
\label{callK}
\end{equation}
In these expressions $\theta$ and $\varphi$ are respectively the polar and azimuthal spherical angles describing the dipole vector ${\bf d}$, while $K_{n}(x)$ is the modified Bessel function of the second kind and order $n$.
In addition, $A_{0}(\varphi,\theta) =(3/2)\left[1+\left(3 \cos^2(\varphi)- 2\right)\sin^{2}(\theta)\right]$ and $A_{2}(\varphi,\theta) = (3/2)\left[1-\cos^{2}(\varphi)\sin^{2}(\theta)\right]$.
It is convenient to 
introduce the dimensionless variables $V=v/c$, $Z=z_{a}\omega_{\rm sp}/c$, $\xi = \omega/\omega_{\rm sp}=\nu V/Z$, 
and $w=k_{x}z_{a}$, where $\omega_{\rm sp}$ represents a characteristic frequency of the surface 
(e.g., the surface plasmon frequency in the case of a metallic medium).
We also define two other auxiliary functions,
\begin{equation}
G^{\theta}\left(\xi,\frac{V}{Z}\right) \equiv \int_{-\infty}^{\infty}\frac{dw}{2\pi}\theta\left(\xi+w\frac{V}{Z}\right)g\left(w,\xi+w\frac{V}{Z}\right) ~,
\end{equation} 
and $G\left(\xi,V/Z\right)$, where the latter differs from the former only by the absence of the Heaviside 
theta function in the integrand.
Using these definitions, the complex atomic polarizability in equation \eqref{alpha} can be rewritten as 
\begin{equation}
 \alpha\left(\xi,Z,\frac{V}{Z}\right) =\frac{\alpha_{\rm sp}}{1-\left(\frac{\xi}{\xi_{a}}\right)^{2}-\frac{\alpha_{\rm sp}}{Z^{3}} G\left(\xi,\frac{V}{Z}\right)}~,
\end{equation}
where  $\xi_{a}=\omega_{a}/\omega_{\rm sp}$ and $\alpha_{\rm sp}=(3\alpha_{0}/\epsilon_{0})\left(\omega_{\rm sp}/c\right)^{3}$. 
These expressions and the power spectrum given in equation \eqref{spectrum} allow us to recast the quantum
frictional force as
\begin{widetext}
\begin{equation}
\frac{F_{\rm fric}}{F_{0}}=-\frac{4}{\alpha_{\rm sp}} \frac{V}{Z^{8}}\int_{0}^{\infty}d\nu\int_{-\infty}^{\infty}\frac{dw}{2\pi} w \abs{\alpha\left([w-\nu]\frac{V}{Z},Z,\frac{V}{Z}\right)}^{2}
g_{I}\left(w,\nu\frac{V}{Z}\right)
 G^{\theta}_{I}\left([w-\nu]\frac{V}{Z},\frac{V}{Z}\right).
\label{frictionDL}
\end{equation} 
\end{widetext}
where $F_{0}= -3\hbar \omega_{\rm sp}^5 \alpha_{0} / (2 \pi \epsilon_{0}c^4)$.

We are now in a position to analyze the low-velocity limit, which corresponds to $V/Z \ll 1$. Assuming 
an Ohmic surface, for small $\xi$ we have $r(\xi)\approx r_{0}+\imath \eta\, \xi$, where $r_0$ is the 
static reflection coefficient. The prefactor $\eta$ is related to the material resistivity $\rho$ by
$\eta=2\rho \epsilon_{0}\omega_{\rm sp}$ (for a metallic surface 
$\eta=\Gamma/\omega_{\rm sp}$). 
The Heaviside function and the modified Bessel functions in the integrand of equation \eqref{frictionDL}
limit the integration range of $w$ and $\nu$, and in the limit $V/Z \ll 1$, we can approximate
%\begin{subequations}
\begin{equation}
 g_{I}\left(w,\nu\frac{V}{Z}\right) \approx
\nu\frac{V}{Z} \frac{\eta}{6\pi}w^{2}\mathcal{K}(w,\varphi,\theta),
\end{equation}
and 
\begin{multline}
 G^{\theta}_{I}\left([w-\nu]\frac{V}{Z},\frac{V}{Z}\right) \approx
\theta(w-\nu)\frac{V}{Z}\eta\frac{w-\nu}{24\pi}\mathcal{A}^{0}(\varphi,\theta)\\
+\frac{V}{Z}\eta\int_{\abs{w-\nu}}^{\infty}\frac{dw_{1}}{2\pi}\frac{(w_{1}-\abs{w-\nu})}{6\pi}w_{1}^{2}\mathcal{K}(w_{1},\varphi,\theta) .
\label{GthetaSplit}
\end{multline}
%\end{subequations}
Here, we have defined $\mathcal{A}^{0}=(A_{0}+3A_{2})/4$ (where $A_0$ and $A_2$ were already defined after 
equation \eqref{callK})
and, for simplicity, have dropped the
arguments. In the same $V/Z \ll 1$ limit, we approximate the polarizability by its $V=0$ value, i.e.,
$\alpha\left([w-\nu]V/Z,Z,V/Z\right)
 \approx 
 \alpha_{\rm sp} \left[1-(\alpha_{\rm sp}r_{0}\mathcal{A}^{0})/(24\pi Z^{3}) \right]^{-1}$. 
From these expressions we can already conclude that for low velocities $F_{\rm fric} \simeq V^3/Z^{10}$.

The two terms on the right-hand side of equation \eqref{GthetaSplit} are related to the two contributions 
to the non-equilibrium fluctuation-dissipation theorem, equation \eqref{NEFDT}. The first term gives the 
low-velocity behavior of $F^{\rm LTE}_{\rm fric}$, and for a metallic surface it can be written as
\begin{equation}
\frac{F^{\rm LTE}_{\rm fric}}{F_{0}}
\approx
\frac{45 \mathcal{A}^{\rm LTE}}{16}
\frac{\frac{\alpha_{0}}{8\pi \epsilon_{0} z_{a}^{3}}\frac{\Gamma^{2}}{24\pi \omega_{\rm sp}^{2}}\left(\frac{c}{\omega_{\rm sp}z_a}\right)^7
\left(\frac{v}{c} \right)^3}{\abs{1-\frac{\alpha_{0}r_{0}\mathcal{A}^{0}}{8 \pi \epsilon_{0} z_{a}^{3}}}^{2}}~,
\label{falpha}
\end{equation} 
where we have defined $\mathcal{A}^{\rm LTE}=(A_{0}+3A_{2})(5 A_{0}+7 A_{2})/48$. Upon neglecting 
the frequency shift in the denominator of the polarizability and averaging over all dipole orientations, we recover the
expression given in equation \eqref{fricalpha}.

The second term in \eqref{GthetaSplit} gives the low-velocity expansion of $F^{J}_{\rm fric}$. 
Its evaluation is more involved than the previous one, and we will only sketch the main steps. 
From the second term in \eqref{GthetaSplit} we can write
\begin{multline}
F^{J}_{\rm fric}
\approx -4
\frac{\hbar \omega_{\rm sp}}{\lambda_{\rm sp}} \frac{V^{3}}{Z^{10}}\abs{\frac{\alpha_{\rm sp}}{1-\frac{\alpha_{\rm sp}\mathcal{A}^{0}}{24\pi Z^{3}}}}^{2} \left(\frac{\eta}{6\pi}\right)^{2}
\\
\times
\int_{-\infty}^{\infty}\frac{dw}{2\pi} \, 
w^{3}\mathcal{K}(w,\varphi,\theta) j(w,\varphi,\theta)~, 
\label{fjoriginal}
\end{multline}
where we have defined the function 
\begin{align}
j(w)&=\int_{-\infty}^{\infty}d\nu \nu\int_{\abs{\nu-w}}^{\infty}\frac{dw_{1}}{2\pi}\frac{(w_{1}-\abs{\nu-w})}{2}w_{1}^{2}\mathcal{K}(w_{1},\varphi,\theta)\nonumber\\
&=\frac{3w}{32}\frac{3 A_{0}(\varphi,\theta)+5 A_{2}(\varphi,\theta)}{8}.
\end{align}
Inserting this result back in \eqref{fjoriginal} and integrating over $w$, we obtain an expression 
similar to equation \eqref{falpha} but where the prefactor $45 \mathcal{A}^{\rm LTE}(\varphi,\theta)/16$ is 
replaced by $9 \mathcal{A}^{J}(\varphi,\theta)/4$, with $\mathcal{A}^{J}(\varphi,\theta)=[(3 A_{0}(\varphi,\theta)+5 A_{2}(\varphi,\theta))/8]^{2}$. Proceeding as for the previous expression we obtain equation \eqref{fricj}.

\vspace{0.5cm}

\noindent F. Intravaia$^{1}$, R. O. Behunin$^{2}$, C. Henkel$^{3}$,K. Busch$^{1,4}$ and D. A. R. Dalvit$^{5}$.
%\vspace{0.5cm}
\begin{small}
\begin{enumerate}
%\addtolength{\itemsep}{-0.4\baselineskip}
\item[$^{1}$]
Max-Born-Institut, 12489 Berlin, Germany
\item[$^{2}$]
Department of Applied Physics, Yale University, New Haven, Connecticut 06511, USA
\item[$^{3}$]
Institute of Physics and Astronomy, University of Potsdam, Karl-Liebknecht-Str. 24/25, 14476 Potsdam, Germany
\item[$^{4}$]
Humboldt-Universit\"at zu Berlin, Institut f\"ur Physik, AG Theoretische Optik \& Photonik, 12489 Berlin, Germany
\item[$^{1}$]
Theoretical Division, MS B213, Los Alamos National Laboratory, Los Alamos, New Mexico 87545, USA
\end{enumerate}
\end{small}

%\vspace{1cm}
%\noindent {\bf Acknowledgments.}
%We acknowledge support by the LANL LDRD program, and by the Deutsche Forschungsgemeinschaft (DFG) through 
%project B10 within the Collaborative Research Center (CRC) 951 Hybrid Inorganic/Organic Systems 
%for Opto-Electronics (HIOS). FI further acknowledges financial support from the European Union Marie Curie People program through the Career Integration Grant No. PCIG14- GA-2013-631571. CH and FI acknowledge support from the DFG through the 
%DIP program (grant FO 703/2-1 and SCHM 1049/7-1)\\

%\noindent {\bf  Competing Interests.} The author declare that they have no
%competing financial interests.\\
%
%\noindent {\bf Correspondence.} Correspondence and requests for materials should be addressed to F.I.~(email: francesco.intravaia@mbi-berlin.de).

%%%%%%%%%%%%%%%%%%%%%%%%%%%%%%%%%%%%%%%%%%%%%%%%%%%%

%\newpage
%\cleardoublepage
%\setcounter{page}{1}
%\setcounter{figure}{0}
%\setcounter{equation}{0}
%\renewcommand{\theequation}{S\arabic{equation}}
%\renewcommand{\figurename}{\textbf{Supplementary Figure}}
%\renewcommand{\thefigure}{{\bf S\arabic{figure}}}
%
%\noindent{\bf {\large Vacuum Incalescence -- SUPPLEMENTARY INFORMATION}\\
%F. Intravaia -- Max-Born-Institut, 12489 Berlin, Germany}
%
%\vspace{1cm}

\end{document}